\documentstyle[12pt,bezier]{article}
\begin{document}
% ***************    NEW COMMANDS   *******************
\def \inbar{\vrule height1.5ex width.4pt depth0pt}
\def \xC{\relax\hbox{\kern.25em$\inbar\kern-.3em{\rm C}$}}
\def \xR{\relax{\rm I\kern-.18em R}}
\newcommand{\xZ}{Z \hspace{-.08in}Z}
\newcommand{\xbe}{\begin{equation}}
\newcommand{\xee}{\end{equation}}
\newcommand{\xbea}{\begin{eqnarray}}
\newcommand{\xeea}{\end{eqnarray}}
\newcommand{\xnn}{\nonumber}
\newcommand{\xkt}{\rangle}
\newcommand{\xbr}{\langle}
\newcommand{\xlll}{\left( }
\newcommand{\xrrr}{\right)}
\newcommand{\xcun}{\mbox{\footnotesize${\cal N}$}}
\newcommand{\cun}{\mbox{\footnotesize${\cal N}$}}
\title{Generalized Adiabatic Product Expansion: 
A~nonperturbative~method of solving time-dependent Schr\"odinger 
equation}
\author{Ali Mostafazadeh\thanks{E-mail address: 
amostafazadeh@ku.edu.tr}\\ \\
Department of Mathematics, Ko\c{c} University,\\
Istinye 80860, Istanbul, TURKEY}
\date{ }
\maketitle

\begin{abstract}
We outline a method based on successive canonical transformations which
yields a product expansion for the evolution operator of a general (possibly
non-Hermitian) Hamiltonian. For a class of such Hamiltonians this 
expansion involves a finite number of terms, and our method gives the 
exact solution of the corresponding time-dependent Schr\"odinger equation.
We apply this method to study the dynamics of a general nondegenerate 
two-level quantum system, a time-dependent classical harmonic oscillator, 
and a degenerate system consisting of a spin 1 particle interacting with a 
time-dependent electric field $\vec{\cal E}(t)$ through the Stark Hamiltonian 
$H=\lambda(\vec J\cdot \vec{\cal E})^2$. 
\end{abstract}
%\vspace{3mm}
%PACS numbers: 03.65.Bz
%\vspace{3mm}

%\baselineskip=30pt

\section*{I. Introduction}
Recently, a method based on successive canonical transformations has been 
used to obtain exact solution of the Schr\"odinger equation	
	\xbe
	i\frac{d}{dt}|\psi(t)\xkt=H(t)|\psi(t)\xkt
	\label{sch-eq}
	\xee
for a class of dipole Hamiltonians \cite{pla97-1,pra97-1,jmp97-2} and
time-dependent harmonic oscillators \cite{pra97-2}. For these systems the 
Hamiltonian is a nondegenrate Hermitian operator. The purpose of the 
present article is to extend the application of this method to the cases where
the Hamiltonian is non-Hermitian and involves degenerate eigenvalues.

Non-Hermitian Hamiltonians have been used to model a variety of physical
systems involving decaying states, \cite{nhh-application}. The solution
of the Schr\"odinger equation for a time-dependent two-level non-Hermitian
Hamiltonian has been considered in Refs.~\cite{da-to-mi,kv-pu}. Another 
motivations for the study of the Schr\"odinger equation for a 
time-dependent non-Hermitian Hamiltonian is the fact that the solution of 
every linear ordinary differential equation (ODE) may be reduced to the 
solution of a system of first order linear ODEs which can be written in the 
form of the time-dependent Schr\"odinger equation~(\ref{sch-eq}) or 
alternatively
	\xbea
	|\psi(t)\xkt&=&U(t)|\psi(0)\xkt\;,
	\label{psi=}\\
	i\frac{d}{dt}U(t)&=&H(t)U(t)\;,
	\label{sch-eq-u}\\
	U(t)&=&1\;,
	\label{ini-condi}
	\xeea
where $U(t)$ is the evolution operator. For a general linear ODE 
the corresponding Hamiltonian $H(t)$ may be a non-Hermitian matrix with 
degenerate eigenvalues. 

The method of {\em adiabatic product expansion} developed in 
Refs.~\cite{pla97-1,pra97-1} does not directly apply to quantum systems
with non-Hermitian Hamiltonians. In this article  we shall present a 
generalization of this method which applies to arbitrary (possibly) 
non-Hermitian Hamiltonians with degenerate as well as nondegenerate 
eigenvalues. 

The organization of the article is as follows. In section~II we review the
basic results concerning the adiabatic approximation for degenerate and
non-Hermitian Hamiltonians. In section~III we discuss the generalization of 
the method of adiabatic product expansion to these Hamiltonians. In 
section~IV we use the results of section~III to study the solution of the 
Schr\"odinger equation for a general nondegenerate non-Hermitian two-level 
Hamiltonian. In section~V, we apply the general results obtained in
section~IV to treat the classical equation of motion for a harmonic 
oscillator with a time-dependent frequency. In section~VI, we discuss the 
application of the adiabatic product expansion to study the quadrupole 
interaction of a spin 1 particle with a time-dependent electric field 
$\vec{\cal E} =({\cal E}_1(t), {\cal E}_2(t),0)$. We show that the 
corresponding Hamiltonian which has a degenerate and a nondegenerate 
eigenvalue is canonically equivalent to a Hamiltonian which has only 
nondegenerate eigenvalues. Furthermore, we show that if the direction of the 
electric field depends in a particular way on its magnitude, then our method 
yields the exact solution of the Schr\"odinger equation. Finally we present our 
conclusions in section~VII.

\section*{II. Adiabatic Approximation for Non-Hermitian Hamiltonians}

Let $H=H[R]$ be a parametric Hamiltonian which depends on a set of real 
parameters $R=(R^1,R^2,\cdots,R^d)$ labelling the points of a smooth 
manifold $M$. Let $E_n[R]$ denote the eigenvalues of $H[R]$
and ${\cal H}_{n}[R]$ be the degeneracy subspace associated with 
$E_n[R]$. Let $\cun$ denote the degree of degeneracy of $E_n[R]$, i.e., 
the complex dimension of ${\cal H}_{n}[R]$. We shall assume that the 
spectrum of  $H[R]$ is discrete and $\cun$ does not depend on $R$. 

Now let $|\psi_n,a;R\xkt$ and $|\phi_n,a;R\xkt$ form a complete 
biorthonormal basis of the Hilbert space \cite{wong,ga-wr}. This means that 
$|\psi_n,a;R\xkt$ with $a\in\{1,2,\cdots,\cun\}$ form a basis of 
${\cal H}_{n}[R]$, in particular
	\xbe
	H[R]|\psi_n,a;R\xkt=E_n[R]|\psi_n,a;R\xkt\;,
	\label{eg-va-eq-H}
	\xee
and $|\phi_n,a;R\xkt$ satisfy
	\xbea
	&&H[R]^\dagger|\phi_n,a;R\xkt=E_n^*[R]|\phi_n,a;R\xkt\;,
	\label{eg-va-eq-H*}\\
	&&\xbr\phi_m,b;R|\psi_n,a;R\xkt=\delta_{mn}\delta_{ab}\;,
	\label{biorthonormal}\\
	&&\sum_{n}\sum_{a=1}^{\cun}|\psi_n,a;R\xkt\xbr\phi_n,a;R|=1\;.
	\label{complete}
	\xeea

Next suppose that the parameters $R^i$ depend on time $t$, then $R(t)$ 
defines a curve ${\cal C}$ in the parameter space $M$, and the Hamiltonian, its 
eigenvalues and eigenvectors become time-dependent. In this case we use 
the notation $H(t):=H[R(t)]$, $E_n(t):=E_n[R(t)]$, $|\psi_n,a;t\xkt:=
|\psi_n,a;R(t)\xkt$, and $|\phi_n,a;t\xkt:=|\phi_n,a;R(t)\xkt$. We shall 
assume that $E_n(t),~|\psi_n,a;t\xkt$ and $|\phi_n,a;t\xkt$ are smooth 
functions of $t$ and that during the evolution of the system the eigenvalues 
of the Hamiltonian do not cross, i.e., if $E_m(0)<E_n(0)$, then for all $t\in
[0,\tau]$, $E_m(t)<E_n(t)$, where $\tau$ denotes the duration of the 
evolution of the system.

Differentiating both sides of Eq.~(\ref{eg-va-eq-H}) with respect to $t$,
taking the inner product of both sides of the resulting equation with
$|\phi_m,b;t\xkt$, for arbitrary $m$ and $b$, and using 
Eqs.~(\ref{eg-va-eq-H}) -- (\ref{biorthonormal}), we have
	\xbe
	[E_m(t)-E_n(t)]\xbr\phi_m,b;t|\frac{d}{dt}|\psi_n,a;t\xkt+
	\xbr\phi_m,b;t|\dot H(t)|\psi_n,a;t\xkt
	-\delta_{mn}\delta_{ab}\dot E(t)=0\;.
	\label{1}
	\xee
Here a dot denotes differentiation with respect to $t$. For $m\neq n$,
Eq.~(\ref{1}) reads
	\xbe
	\xbr\phi_m,b;t|\frac{d}{dt}|\psi_n,a;t\xkt=\frac{
	\xbr\phi_m,b;t|\dot H(t)|\psi_n,a;t\xkt}{E_n(t)-E_m(t)}~~~
	{\rm for}~~~m\neq n.
	\label{2}
	\xee

Now let us express the solution of the Schr\"odinger equation~(\ref{sch-eq})
in the basis $\{|\psi_n,a;t\xkt\}$. Then
	\xbe
	|\psi(t)\xkt=\sum_{n}\sum_{a=1}^{\cun}C_a^n(t)|\psi_n,a;t\xkt\;,
	\label{3}	
	\xee
where $C_a^n(t)$ are complex coefficients. Substituting Eq.~(\ref{3}) in the 
Schr\"odinger equation~(\ref{sch-eq}), taking the inner product of both sides 
of the resulting equation with $|\phi_m,b;t\xkt$, and making use of 
Eqs.~(\ref{eg-va-eq-H}), (\ref{eg-va-eq-H*}), (\ref{biorthonormal}),
and (\ref{2}), we find
	\xbe
	i\dot C_b^m-E_m C_b^m+\sum_{a=1}^{\cun} i\xbr\phi_m,b;t|\frac{d}{dt}
	\psi_m,a;t\xkt C_a^m=-i\sum_{n\neq m}\sum_{a=1}^{\cun}
	\frac{\xbr\phi_m,b;t|\dot H(t)|\psi_n,a;t\xkt}{E_n(t)-E_m(t)}\;.
	\label{4}
	\xee
The special case of this equation with $\cun=1$, i.e., the nondegenerate case,
has been originally derived by Garrison and Wright \cite{ga-wr} in their 
investigation of the adiabatic geometric phase \cite{berry1984} for 
non-Hermitian Hamiltonians \cite{ga-wr,da-mi-to,mi-si-ba-be,mo-he,p28}.

If the right-hand side of Eq.~(\ref{4}) is negligible, then one says that the
system undergoes an {\rm adiabatic evolution}, 
\cite{bo-fo,kato,pra97-1,ga-wr,ne-ra}. In this
case, the equations for  $C_a^n$ decouple and their solution is given by	
	\xbe
	C^n_a(t)=\sum_{b=1}^{\cun} K^n_{ab}(t)C^n_b(0)\;,
	\label{5}
	\xee
where $K_{ab}^n(t)$ are entries of the invertible matrix
	\xbe
	K^n(t):=e^{-i\int_0^tE_n(s)ds}\;{\cal P}
	\exp \left[i\int_{R(0)}^{R(t)} {\cal A}^n[R]\right]\;,
	\label{6}
	\xee
${\cal P}$ denotes the path-ordering operator, ${\cal A}^n$
is the matrix of one-forms with entries
	\xbe
	{\cal A}^n_{ab}[R]:=i\xbr\phi_n,a;R|d|\psi_n,b;R\xkt\;,
	\label{a=}
	\xee
$d$ stands for the exterior derivative with respect to $R^i$, and
the line intergral in Eq.~(\ref{6}) is evaluated along the curve 
${\cal C}$ defined by $R(t)$. If ${\cal C}$ is a closed curve in $M$, the 
Hamiltonian has a periodic time-dependence and the path-ordered exponential
in Eq.~(\ref{6}), which takes the form
	\xbe
	{\cal P}\exp \left[i\oint_{\cal C}{\cal A}^n[R]\right]\;,
	\label{gp}
	\xee	
is the {\em non-Hermitian} analogue of the {\em non-Abelian adiabatic
geometric phase} \cite{wi-ze}.

Note that if the initial vector $|\psi(0)\xkt$ is an eigenvector of the
initial Hamiltonian $H(0)$, then the adiabaticity of the evolution implies
that $|\psi(t)\xkt$ is an eigenvector of $H(t)$ for all $t\in[0,\tau]$. In terms 
of the time-evolution operator $U(t)$ of Eq.~(\ref{sch-eq-u}) this is 
expressed by
	\xbea
	U(t)&\approx& U^{(0)}(t)\;,~~~{\rm where}
	\label{7}\\
	U^{(0)}(t)&:=&\sum_{n}\sum_{a,b=1}^{\cun} K^n_{ab}(t)
	|\psi_n,a;t\xkt\xbr\phi_n,b;0|\;.
	\label{u0}
	\xeea

One can easily show that $U^{(0)}(t)$ is invertible, and its inverse is
given by
	\xbe
	U^{(0)^{-1}}(t)=\sum_n\sum_{a,b=1}^{\cun} K^{n^{-1}}_{ab}(t)
	|\psi_n,a;0\xkt\xbr\phi_n,b;t|\;,
	\label{u0-1}
	\xee
where $K^{n^{-1}}(t)$ is the inverse of $K^n(t)$.

\section*{III. Adiabatic Canonical Transformations and the Generalized
Adiabatic Product Expansion}

Let $g(t)$ be an invertible linear operator acting on the Hilbert space.
Then the transformations:
	\xbea
	|\psi(t)\xkt&\to& |\psi'(t)\xkt:=g(t)|\psi(t)\xkt\;,
	\label{psi-trans}\\
	H(t)&\to& H'(t):=g(t)H(t)g(t)^{-1}-ig(t)\frac{d}{dt}g(t)^{-1}\;,
	\label{H-trans}\\
	U(t)&\to&U'(t):=g(t)U(t)g(0)^{-1}\;,
	\label{u-trans}
	\xeea
leave the form of the Schr\"odinger equation invariant. We shall call
such a transformation a {\em canonical transformation.}

Now let us investigate the consequences of the canonical transformation
defined by $g(t)=U^{(0)}(t)^{-1}$. We shall call this transformation the 
{\em adiabatic canonical transformation.} Denoting the transformed 
Hamiltonian $H'$ by $H^{(1)}$, we have
	\xbea
	H^{(1)}(t)&=&\sum_{n,m\neq n}\sum_{a=1}^{\cun}
	\sum_{b=1}^{\cal M}H^{(1)^{nm}}_{ab}(t)|\psi_n,a;0\xkt\xbr
	\phi_m,b;0|\;,~~~{\rm where}
	\label{H1}\\
	H^{(1){nm}}_{ab}(t)&:=&-K_{ac}^{n}(t)^{-1}A_{cd}^{nm}(t)
	K_{db}^m(t)\;,~~~{\rm and}~~~
	A_{cd}^{nm}(t):=i\xbr\phi_n,c;t|\frac{d}{dt}|\psi_m,d;t\xkt\;.
	\label{H1-ab}
	\xeea
Because $g(0)=U^{(0)}(0)^{-1}=1$, the transformed evolution operator is
given by
	\xbe
	U'(t)=U^{(0)}(t)^{-1} U(t) \;.
	\label{u'=}
	\xee
Clearly if the adiabatic approximation is valid, $H^{(1)}(t)\approx 0$ 
and $U'(t)\approx 1$.

Let us suppose that $H^{(1)}(t)$ has a discrete spectrum and denote
by $E^{(1)}_{n_1}(t)$ and $\cun_1$ the eigenvalues of $H^{(1)}(t)$ and
their degree of degeneracy. Furthermore, let $\{|\psi_{n_1}^{(1)},a_1;t\xkt,
|\phi_n^{(1)},a_1;t\xkt\}$ be a biorthonormal eigenbasis of the Hilbert
space, i.e.,
	\xbea
	&&H^{(1)}(t)|\psi^{(1)}_{n_1},a_1;t\xkt=E^{(1)}_{n_1}[R]
	|\psi^{(1)}_{n_1},a_1;t\xkt\;,\xnn\\
	&&H^{(1)}(t)^\dagger|\phi^{(1)}_{n_1},a_1;t\xkt=E^{(1)*}_{n_1}(t)
	|\phi^{(1)}_{n_1},a_1;t\xkt\;,\xnn\\
	&&\xbr\phi^{(1)}_{m_1},b_1;t|\psi^{(1)}_{n_1},a_1;t\xkt=
	\delta_{m_1n_1}\delta_{a_1b_1}\;,\xnn\\
	&&\sum_{n_1}\sum_{a_1=1}^{\cun_1}|\psi^{(1)}_{n_1},a_1;t\xkt
	\xbr\phi^{(1)}_{n_1},a_1;t|=1\;.\xnn
	\xeea
Then $H^{(1)}(t)$ shares the properties of the original Hamiltonian $H(t)$, 
and we can repeat the above analysis using $H^{(1)}(t)$ in place of $H(t)$. 
In this way the adiabatic approximation yields the approximate evolution 
operator
	\xbe
	U^{(1)}(t)=\sum_{n_1}\sum_{a_1,b_1=1}^{\cun_1} 
	K^{(1)n_1}_{a_1b_1}(t)|\psi_{n_1}^{(1)},a_1;t\xkt
	\xbr\phi_{n_1}^{(1)},b_1;0|
	\label{u1=}
	\xee
for $H^{(1)}(t)$, where $K^{(1)n_1}_{a_1b_1}(t)$ are the entries of the 
matrix $K^{(1)n_1}$ obtained by replacing $E_n(t),~ |\psi_n,a;t\xkt$ and
$|\phi_n,a;t\xkt$ in Eqs.~(\ref{6}) and (\ref{a=}) by $E_{n_1}^{(1)}(t)$,
$|\psi_{n_1}^{(1)},a_1;t\xkt$, and $|\phi_{n_1}^{(1)},a_1;t\xkt$, 
respectively.

Next we perform the adiabatic canonical transformation defined by $g(t)=
U^{(1)}(t)^{-1}$. This leads to a transformed Hamiltonian $H^{(2)}(t)$ 
which is related to $H^{(1)}(t)$ according to Eqs.~(\ref{H1}) and
(\ref{H1-ab}) with $K^n,~|\psi_n,a;t\xkt$ and $|\phi_n,b;t\xkt$ replaced 
by $K^{(1)^{n_1}},~|\psi_{n_1}^{(1)},a_1;t\xkt$ and 
$|\phi_{n_1}^{(1)},a_1;t\xkt$. The transformed evolution operator is given
by
	\[U^{(1)}(t)^{-1}U^{(0)}(t)^{-1}U(t)\;.\]

Repeating this procedure we obtain, after $N$ successive adiabatic canonical 
transformations, a transformed Hamiltonian $H^{(N)}(t)$ and a transformed 
evolution operator which is given by
	\[U^{(N-1)}(t)^{-1}U^{(N-2)}(t)^{-1}\cdots U^{(0)}(t)^{-1}U(t)\;.\]
Here $U^{(\ell)}(t)$, with $\ell\in\{1,2,\cdots,N-1\}$, denotes the approximate 
evolution operator obtained by performing adiabatic approximation on the 
Hamiltonian $H^{(\ell)}(t)$.

If for some $N$ the adiabatic approximation yields the exact solution
of the Schr\"odinger equation for the Hamiltonian $H^{(N)}(t)$, then by
construction $H^{(N+1)}(t)=0$ and $U^{(N+1)}(t)=1$. In this case, the
original evolution operator is given by
	\xbe
	U(t)=U^{(0)}(t)U^{(1)}(t)\cdots U^{(N)}(t)\;.
	\label{u=approx}
	\xee
If the adiabatic approximation fails for all $H^{(N)}(t)$, then there are
two possibilities:
	\begin{itemize}
	\item[i)] one obtains an infinite product expansion for the evolution
	operator
	\xbe
	U(t)=\prod_{\ell=0}^{\infty} U^{(\ell)}(t):=
	U^{(0)}(t)U^{(1)}(t)\cdots U^{(\ell)}(t)\cdots\;.
	\label{u=exact}
	\xee
	In this case, one may view Eq.~(\ref{u=approx}) as a {\em 
	generalization} of  the adiabatic approximation.
	\item[ii)] one obtains $H^{(i)}(t)=H^{(j)}(t)$ for some $i$ and $j$ 
	with $i\neq j$.
	In this case a direct application of the method of adiabatic product 
	expansion  does not produce a solution. However, as we shall see in 
	the following section, sometimes it is possible to modify this method 
	by combining the adiabatic canonical transformation with other
	canonical transformations, so that one obtains a finite or an infinite
	product expansion with distinct terms.
	\end{itemize}

\section*{IV. Application to Two-Level Hamiltonians}

Two-level nondegenerate Hamiltonians provide the simplest nontrivial 
quantum systems. This has been one of the main reasons for the study of
these Hamiltonians since the early days of quantum mechanics. In this 
section we shall consider the most general nondegenerate two-level 
Hamiltonian which may or may not be Hermitian. 

In an arbitrary basis of the Hilbert space ($\xC^2$), the Hamiltonian 
is given by a two-by-two complex matrix $\bar H$. One can perform a quantum 
canonical transformation (\ref{H-trans}) defined by $g(t)=\exp\{i\int_0^t 
[{\rm tr}~\bar H(s)] ds/2\}$ to map the Hamiltonian $\bar H$ to a traceless 
Hamiltonian of the form
	\xbe
	H:=\left(\begin{array}{cc}a & b \\
	c&-a
	\end{array}\right)\;,
	\label{2-level-H-traceless}
	\xee
where ${\rm tr}~ \bar H$ denotes the trace of $\bar H$, and $a=a(t),~b=b(t),~c=c(t)$ 
are complex-valued smooth functions of $t$.

We can easily solve the eigenvalue problem for the Hamiltonian (\ref{2-level-H-traceless}). The 
eigenvalues are given by
	\xbe
	E_1(t):=-E(t),~~~E_2(t):=E(t)\;,~~~{\rm where}~~~E:=\sqrt{a^2+bc}. 
	\label{2-level-eg-va}
	\xee
We shall demand that during the time interval $[0,\tau]$ of interest $E\neq 0$, 
so that the eigenvalues are nondegenerate. In particular, no level 
crossings occur. Then a possible choice for a biorthonormal eigenbasis is
	\xbea
	&&|\psi_1;R\xkt=\left(\begin{array}{c}
	-b\\
	a+E\end{array}\right)\,,~~~
	|\psi_2;R\xkt=\left(\begin{array}{c}
	a+E\\
	c\end{array}\right)\,
	\label{eg-vec}\\
	&&|\phi_1;R\xkt=\frac{1}{N^*}\left(\begin{array}{c}
	-c^*\\
	a^*+E^*\end{array}\right)\,,~~~
	|\phi_2;R\xkt=\frac{1}{N^*}\left(\begin{array}{c}
	a^*+E^*\\
	b*\end{array}\right)\;,
	\label{dual-eg-vec}
	\xeea
where $R=(a,b,c)$ and $N:=2E(a+E)$.	

Next we compute $U^{(0)}$ and $H^{(1)}$. Using Eqs.~(\ref{u0}),
(\ref{6}), (\ref{a=}), and ({\ref{H1}), we find
	\xbea
	U^{(0)}(t)&=&K^1(t)|\psi_1;t\xkt\xbr\phi_1;0|+
	K^2(t)|\psi_2;t\xkt\xbr\phi_2;0|\;,
	\label{u0-2}\\
	H^{(1)}(t)&=&\xi(t)|\psi_1;0\xkt\xbr\phi_2;0|+\zeta(t)
	|\psi_2;0\xkt\xbr\phi_1;0|\;,
	\label{H1-2}
	\xeea
where
	\xbea
	K^1(t)&:=&K^1_{11}(t)=\exp\left( \frac{i\eta(t)}{2}-
	\int_{R(0)}^{R(t)}\left[(2E)^{-1}(da+dE+\frac{cdb}{a+E})\right]
	\right)\;,\xnn\\
	K^2(t)&:=&K^2_{11}(t)=\exp\left( \frac{-i\eta(t)}{2}-
	\int_{R(0)}^{R(t)}\left[(2E)^{-1}(da+dE+\frac{bdc}{a+E})\right]
	\right)\;,\xnn\\
	\eta(t)&:=&2\int_0^t E(s)ds\;,
	\label{eta}\\
	\xi(t)&:=&H^{12}_{11}(t)=(-\frac{ie^{-2i\alpha(t)}}{2})
	\left[1+\frac{a(t)}{E(t)}\right]
	\frac{d}{dt}\left[\frac{c(t)}{a(t)+E(t)}\right]\;,
	\label{xi}\\
	\zeta(t)&:=&H^{21}_{11}(t)=(\frac{ie^{2i\alpha(t)}}{2})
	\left[1+\frac{a(t)}{E(t)}\right]
	\frac{d}{dt}\left[\frac{b(t)}{a(t)+E(t)}\right]\;,
	\label{zeta}\\
	\alpha(t)&:=&\frac{\eta(t)}{2}+\frac{i}{4}\int_{R(0)}^{R(t)}\frac{
	cdb-bdc}{E(E+a)}\;.
	\label{alpha}
	\xeea

The transformed Hamiltonian has the following matrix expression
	\xbe
	H^{(1)}(t)=\left(\begin{array}{cc} a^{(1)}(t) & b^{(1)}(t)\\
	c^{(1)}(t) & -a^{(1)}(t)\end{array}\right)\;,
	\label{H1-2-matrix}
	\xee
where 
	\xbea
	a^{(1)}(t)&:=&-\frac{b_0\xi(t)+c_0\zeta(t)}{2E_0}\;,
	\label{a1}\\
	b^{(1)}(t)&:=&-\frac{b_0^2\xi(t)-(a_0+E_0)^2\zeta(t)}{2E_0(E_0+a_0)}\;,
	\label{b1}\\
	c^{(1)}(t)&:=&-\frac{-(a_0+E_0)^2\xi(t)+c_0^2\zeta(t)}{2E_0(E_0+a_0)}\;,
	\label{c1}\\
	a_0&:=&a(0)\;,~~~b_0:=b(0)\;,~~~c_0:=c(0)\;,~~~E_0:=E(0)\;,\xnn
	\xeea
and we have used Eqs.~(\ref{eg-vec}), (\ref{dual-eg-vec}), and (\ref{H1-2}).

Note that the transformed Hamiltonian $H^{(1)}(t)$  is traceless, and
one can obtain $H^{(2)}(t)$ by substituting $a^{(1)}$ for $a$,
$b^{(1)}$ for $b$, $c^{(1)}$ for $c$, and $E^{(1)}:=\sqrt{(a^{(1)})^2+
b^{(1)}c^{(1)}}$ for $E$ in Eqs.~(\ref{H1-2}), and (\ref{xi}) --
(\ref{alpha}). Clearly this can be repeated indefinitely, and one
can compute $H^{(\ell)}$ for arbitrary $\ell$.

The adiabatic approximation corresponds to the cases where the matrix
elements of $H^{(1)}(t)$ can be neglected. As seen from 
Eqs.~(\ref{H1-2-matrix}) -- (\ref{c1}) this happens whenever both $\xi$ 
and $\zeta$ are negligible. One can also check that if only one of these 
quantities is negligible, then $H^{(1)}(t)$ is equal to the other times a constant 
matrix. This means that $H^{(1)}(t)$ has essentially stationary eigenvectors 
and the adiabatic approximation would yield the solution of the Schr\"odinger 
equation for $H^{(1)}$. In fact, it is not difficult to check that for the cases 
that either $\xi$ or $\zeta$ is negligible, $H^{(2)}(t)\approx 0$. In particular, 
setting $\xi=0$ or $\zeta=0$ implies $H^{(2)}(t)= 0$ and the evolution 
operator is given by
	\xbe
	U(t)=U^{(0)}(t)U^{(1)}(t)\;.
	\label{u=uu}
	\xee
Therefore, the conditions $\xi=0$ and $\zeta=0$ each define a class of 
exactly solvable two-level systems. In view of Eqs.~(\ref{xi}) and 
(\ref{zeta}), these are
	\begin{itemize}
	\item[] Class 1: The two level systems for which $\frac{c}{a+E}=\mu
	=$ constant, or alternatively $c=\mu(\mu b+\sqrt{4a^2+\mu^2 b^2})/2$;
	\item[] Class 2: The two level systems for which $\frac{b}{a+E}=\nu=$ 
	constant, or alternatively $c=b/\nu^2-a^2/b$.
	\end{itemize}

In general $\xi$ and $\zeta$ do not vanish and the adiabatic product expansion
does not terminate. There is also a special class of two-level systems for which the
product expansion has a periodic structure in the sense of case (ii) of the preceding
section. This is
	\begin{itemize}
	\item[] Class 3: The two level systems for which $a=0$.
	\end{itemize}
Setting $a=0$ in Eqs.~(\ref{alpha}),~(\ref{eta}), (\ref{xi}), and (\ref{zeta}) and
defining $f(t):=i\sqrt{c(t)/b(t)}$, we have
	\xbea
	\alpha(t)&=&\frac{\eta(t)}{2}+\frac{i}{4}\ln\left(\frac{c_0b(t)}{
	b_0c(t)}\right)\;,~~~~
	\eta(t)=2\int_0^t\sqrt{b(s)c(s)}ds\xnn\\
	\xi(t)&=& -\frac{f_0\dot f(t) e^{-i\eta(t)}}{2f(t)}\;,~~~~
	\zeta(t)=\frac{\dot f(t) e^{i\eta(t)}}{2f_0f(t)}\;,\xnn
	\xeea
where $f_0:=f(0)$. Substituting these equations in Eq.~(\ref{H1-2-matrix}),
we obtain
	\xbe
	H^{(1)}(t)=E^{(1)}(t) \left(\begin{array}{cc}
	\cos\eta(t)&f_0^{-1}\sin\eta(t)\\
	f_0\sin\eta(t)&-\cos\eta(t)\end{array}\right)\;,~~{\rm where}~~
	E^{(1)}(t)=\frac{i\dot f(t)}{2f(t)}\;.
	\label{H1-a=0}
	\xee
This Hamiltonian has two interesting properties. 
	\begin{itemize}
	\item[1.] If $b_0=c_0$, then $f_0=i$ and
	\xbe
	H^{(1)}(t) =E^{(1)}(t) \left[ \sin\eta(t)\sigma_2+
	\cos\eta(t)\sigma_3\right]=E^{(1)}(t) e^{i\eta(t)\sigma_1/2}
	\sigma_3e^{-i\eta(t)\sigma_1/2}\;,
	\label{pauli}
	\xee
	where $\sigma_i$ are Pauli matrices, and we have used the
	identity
	\xbe
	e^{-i\varphi\sigma_i}\sigma_je^{i\varphi\sigma_i}=
	\cos(2\varphi)\sigma_j+\sin(2\varphi)\sum_{k=1}^3\epsilon_{ijk}
	\sigma_k\;,~~~~{\rm for}~~~i\neq k\;.
	\label{identity}
	\xee
	In Eq.~(\ref{identity}), $\varphi$ is an arbitrary complex variable
	and $\epsilon_{ijk}$ is the totally anti-symmetric Levi Civita symbol
	with $\epsilon_{123}=1$. For the time periods during which 
	$\sqrt{b(t)c(t)}$ is real, $\eta(t)$ is real, and the Hamiltonian 
	(\ref{pauli}) is anti-Hermitian. In particular, its eigenvectors are 
	orthogonal. Up to a factor of $i$ this Hamiltonian describes the 
	interaction of a spin 1/2 magnetic dipole with a changing magnetic field.
	This system has a $SU(2)$ dynamical group 
	\cite{su2,su,pla97-1,jmp97-2}. For the time periods 
	during which $\sqrt{b(t)c(t)}$ is imaginary, $\eta(t)$ is imaginary, and
	up to a factor of $i$ the Hamiltonian (\ref{pauli}) describes a quantum 
	system with a $SU(1,1)$ dynamical  group. A Hermitian analogue of 
	such a system is the time-dependent generalized harmonic oscillator 
	\cite{su1-1,su,jackiw}.
	\item[2.] Performing the adiabatic canonical transformation on 
	(\ref{pauli}), we arrive at the unexpected result
	\xbe
	H^{(2)}(t)=H(t).
	\label{H2=H}
	\xee
	Therefore, direct application of the method of adiabatic product 
	expansion does not lead to a solution.
	\end{itemize}

Next we shall describe a modification of the method of adiabatic product 
expansion which yields an infinite product expansion for the evolution 
operator of the Class~3 systems which involve distinct terms. 

Consider the transformed Hamiltonian (\ref{H1-a=0}). We can express this 
Hamiltonian using Eq.~(\ref{H1-2-matrix}) with
	\xbe
	a^{(1)}(t)=E^{(1)}(t)\cos\eta(t)\;,~~~
	b^{(1)}(t)=f_0^{-1}E^{(1)}(t)\sin\eta(t)\;,~~~
	c^{(1)}(t)=f_0E^{(1)}(t)\sin\eta(t)\;.
	\label{a1-b1-c1}
	\xee
Although this Hamiltonian does not belong to Class~3, it can be canonically
transformed to a Hamiltonian which belongs to Class~3, i.e., its diagonal 
matrix elements vanish. This transformation is defined by
$g(t)=\exp\{i\int_0^ta^{(1)}(s)ds\sigma_3\} $. The corresponding 
transformed Hamiltonian is given by
	\xbe
	H_1(t)=\left(\begin{array}{cc}
	0 & b_1(t) \\
	c_1(t)  & 0 \end{array}\right) \;,
	\label{H_1}
	\xee
where
	\xbe
	b_1(t):=b^{(1)}(t)e^{i\gamma_1(t)}\;,~~~
	c_1(t):=c^{(1)}(t)e^{-i\gamma_1(t)}\;,~~
	{\rm and}~~\gamma_1 (t):=2\int_0^ta^{(1)}(s)ds\;.
	\label{a_1-b_1-gamma_1}
	\xee
The evolution operator $U_1$ of $H_1$ is related to the evolution operator
of the original Hamiltonian $H$ according to
	\xbe
	U_1(t)=e^{i\int_0^ta^{(1)}(s)ds\sigma_3}~U^{(0)}(t)^\dagger ~U(t)\;,
	\label{u_1}
	\xee
where we have used Eq.~(\ref{u-trans}).

Now since $H_1$ has the same form as $H$, we can repeat the above
analysis using $H_1$ in place of $ H$. Performing an adiabatic canonical 
transformation on $H_1$ we obtain the transformed Hamiltonian 
	\xbe
	H_1^{(1)}(t)=\left(\begin{array}{cc} a_1^{(1)}(t) & b_1^{(1)}(t)\\
	c_1^{(1)}(t) & -a_1^{(1)}(t)\end{array}\right)\;,
	\label{H1-2-matrix-2}
	\xee
where
	\xbea
	a_1^{(1)}(t)&:=&E_1^{(1)}(t)\cos\eta_1(t)\;,~~~
	b_1^{(1)}(t):=f_{1,0}^{-1}E_1^{(1)}(t)\sin\eta_1(t)\;,\xnn\\
	c_1^{(1)}(t)&:=&f_{1,0}E_1^{(1)}(t)\sin\eta_1(t)\;,~~~
	E_1^{(1)}(t):=E^{(1)}(t)\cos\eta(t)\;,\xnn\\
	f_1(t)&:=&i\sqrt{\frac{c_1(t)}{b_1(t)}}=if_0e^{-i\gamma_1(t)}\;,~~~
	f_{1,0}:=f_1(0)=if_0\;,\xnn\\
	\eta_1(t)&:=&2\int_0^tE^{(1)}(s)\sin\eta(s)ds\;.
	\label{eta1}
	\xeea
Clearly we can repeat this procedure indefinitely and construct
an infinite product expansion for the evolution operator. Again if we
compute only a finite number of terms in this expansion, then we obtain
a generalization of the adiabatic approximation. The validity of this
approximation may be checked by computing the transformed 
Hamiltonians. It is not difficult to show that the transformed Hamiltonian 
obtained after $\ell$ adiabatic canonical transformations is of the form
	\[H_\ell^{(1)}(t)= h_\ell(t) S(t)\;,\]
where 
	\xbea
	h_\ell(t)&=&E^{(1)}(t)\cos\eta(t)\cos\eta_1(t)\cos\eta_2(t)
	\cdots\cos\eta_{\ell-1}(t)\;,~~~{\rm and}\xnn\\
	\eta_j(t)&:=&2\int_0^t E^{(1)}(s)\cos\eta(s)\cos\eta_1(s)\cdots
	\cos\eta_{j-1}(s)\sin\eta_{j-1}(s)ds\;,\xnn
	\xeea
where $j\in\{2,3,\cdots,\ell-1\}$ and $S(t)$ is a two-by-two matrix of 
unit determinant. Clearly if for some $\ell$, $h_\ell(t)$ is negligible, then 
the above mentioned generalization of the adiabatic approximation is valid.

Finally let us note that in general the initial Hamiltonian 
(\ref{2-level-H-traceless}) can be written in the form
	\xbe
	H(t)=\alpha_1(t)\sigma_1+\alpha_2(t)\sigma_2+a(t)\sigma_3\;,
	\label{linear-H}
	\xee
with $\alpha_1=(b+c)/2$ and $\alpha_2=i(b-c)/2$. Performing the
canonical transformation (\ref{H-trans}) defined by $g(t)=\exp\{i\int_0^t
\alpha_2(s)ds\sigma_2\}$, we transform the Hamiltonian (\ref{linear-H}) 
into
	\xbea
	H'(t)&=&\alpha'(t)\sigma_1+a'(t)\sigma_3=\left(\begin{array}{cc}
	a'(t) & \alpha'(t)\\
	\alpha'(t) & -a'(t)\end{array}\right)\;,~~~{\rm where}
	\label{linear-H'}\\
	\alpha'(t)&:=&\alpha_1(t)\cos\xi(t)-a(t)\sin\xi(t)\;,~~~~
	a'(t):=\alpha_1(t)\sin\xi(t)+a(t)\cos\xi(t)\;,\xnn\\
	\xi(t)&:=&\int_0^t\alpha_2(s)ds\;.\xnn
	\xeea
Here we have used Eqs.~(\ref{H-trans}) and (\ref{identity}). Next we 
perform another canonical transformation, namely the one defined by $g(t)
=\exp\{i\int_0^t a'(s)ds\sigma_3\}$. This transformation maps the Hamiltonian
(\ref{linear-H'}) into 
	\xbe
	H''(t)=\alpha'(t) e^{i\eta'(t)\sigma_3/2} \sigma_1
	e^{-i\eta'(t)\sigma_3/2}=
	\alpha'(t) [\cos\eta'(t)\sigma_1-\sin\eta'(t)\sigma_2]=\alpha'(t)
	\left(\begin{array}{cc}
	o& e^{i\eta'(t)}\\
	e^{-i\eta'(t)} & 0\end{array}\right)\;,
	\label{linear-H''}
	\xee
where $\eta'(t):=2\int_0^t a'(s)ds$. This Hamiltonian is not only a member of 
Class~3 Hamiltonians, but initially (at $t=0$) its off-diagonal matrix elements 
are equal. In particular, it has the properties 1. in the above list. Note that we
can carry out these canonical transformations on any two-level Hamiltonian.
Therefore, every two-level Hamiltonian is canonically equivalent to a Class~3 
Hamiltonian of the form~(\ref{linear-H''}). This means that the results obtained
for Class~3 Hamiltonians apply to arbitrary two-level Hamiltonians.

\section*{V. Time-dependent Simple Harmonic Oscillator}

It is well-known that the solution of every second order linear ODE 
\cite{ode} can be reduced to the classical equation of motion for a simple 
harmonic oscillator with a time-dependent frequency $\omega=\omega(t)$,
	\xbe
	\ddot x(t)+\omega^2(t) x(t)=0\;.
	\label{sho}
	\xee
It is also well-known that one can reduce
both the classical and quantum equations of motion for a generalized 
harmonic oscillator to Eq.~(\ref{sho}), \cite{hoso,manko,jackiw,jpa98-1}. 
This equation has, therefore, many physical applications 
\cite{manko,application}. Yet an exact analytic expression for the general 
solution of this equation is not known even for the case of real frequency
\cite{belman}. 
\footnote{The lack of an exact analytic solution of  Eq.~(\ref{sho}) is not 
surprising. One way to see this is to recall that the time-independent 
Schr\"odinger equation for an arbitrary potential $V(x)$ in one dimension is 
given by
	\xbe
	\frac{d^2\psi_n}{dx^2}+\left(\frac{\hbar^2[E_n-V(x)]}{2m}\right)\psi_n=0,
	\label{time-indep}
	\xee
where $E_n$ and $\psi_n$ are the energy eigenvalues and eigenfunctions, 
respectively. Eq.~(\ref{time-indep}) can be easily identified with Eq.~(\ref{sho}) 
provided that one makes the change of variables: $x\to t,~\psi_n\to x$,
and $\{\hbar^2[E-V(x)]\}/(2m)\to \omega^2(t)$. This shows that if one was 
able to find the exact analytic solution of  Eq.~(\ref{sho}) for arbitrary 
frequency $\omega$, then one would have been able to find the general 
solution of the time-independent Schr\"odinger equation for any potential 
$V$.}

In the following we shall consider the case of an ordinary time-dependent
harmonic oscillator (\ref{sho}) with real frequency. In order to apply the 
results of the preceding section to Eq.~(\ref{sho}), we first express it in the 
form of a system of first order ODEs. Defining,
	\[ |\psi(t)\xkt:=\left(\begin{array}{c} x(t)\\ v(t)\end{array}\right)\;,~~~
	{\rm and}~~~v(t):=\dot x(t)\;,\]
we can write Eq.~(\ref{sho}) in the form of the Schr\"odinger equation
(\ref{sch-eq}) with a  two-level Hamiltonian of the form 
(\ref{2-level-H-traceless}) with 
%eigenvalues 
%(\ref{2-level-eg-va}) 
%where
	\xbe
	a=0\;,~~~~b=i\;,~~~~c=-i\omega(t)^2\;,~~{\rm and}~~E=\omega(t)\;.
	\label{abc-sho}
	\xee

Since $a=0$, this system belongs to the Class~3 of the preceding section
with 
	\xbea
	f(t)&=&\omega(t)\;,~~~\eta(t)=2\int_0^t\omega(s)ds\;,\xnn\\ 
	H^{(1)}&=&E^{(1)}(t)\left(\begin{array}{cc}
	\cos\eta(t)&\frac{\sin\eta(t)}{\omega_0}\\
	\omega_0\sin\eta(t)&\cos\eta(t)\end{array}\right)\;,~~~
	{\rm and}~~~E^{(1)}(t):=\frac{i\dot\omega(t)}{2\omega(t)}.
	\label{H1-sho}
	\xeea

Clearly, for real frequency $\omega(t)$ we can scale the time variable $t$
so that $\omega_0=1$. Then the transformed Hamiltonian (\ref{H1-sho})
takes the form
	\xbe
	H^{(1)}(t)=E^{(1)}(t) \left[\sin\eta(t)\sigma_1+\cos\eta(t)\sigma_3
	\right]=E^{(1)}(t) e^{-i\eta(t)\sigma_2/2}\sigma_3
	e^{i\eta(t)\sigma_2/2}\;.
	\label{H1-sho-0}
	\xee

Note that since $\eta(t)$ is real, the Hamiltonian (\ref{H1-sho-0}) is an
anti-Hermitian matrix with orthogonal eigenvectors. Therefore, up to a 
factor of $i$ it describes a two-level spin system with a $SU(2)$ 
dynamical group \cite{su2,jmp97-2}.\footnote{Note that one can absorb
the factor of $i$ in the definition of the time variable $t$, i.e., by defining
the imaginary time variable $\tau:=-it$. Therefore, the dynamics given by
the Hamiltonian (\ref{H1-sho-0}) may be viewed as the dynamics of a
spin system with imaginary time.} This is rather surprising, for it is 
well-known that the quantum harmonic oscillator has a $SU(1,1)$ 
dynamical group and that its Schr\"odinger equation may be reduced to 
Eq.~(\ref{sho}) by means of a quantum canonical transformation 
corresponding to a time-dependent dilatation \cite{jpa98-1}.

In view of the fact that $E^{(1)}$ is proportional to the derivative of 
$\ln\omega$, we can make a change of independent variable, namely
$t\to\eta$. Note that  $\eta$ is the integral of a positive real function of $t$. 
Hence, it is a monotonically increasing function of $t$. Making this change
of variable the Schr\"odinger equation for the Hamiltonian (\ref{H1-sho-0})
becomes
	\[i\frac{d}{d\eta}\tilde U(\eta)=\tilde H(\eta)\tilde U(\eta)\;,~~~
	\tilde U(\eta)=1,\]
where $\tilde U(\eta):=U'(t(\eta))$, $U'(t)$ is the evolution operator for the
the Hamiltonian (\ref{H1-sho-0}),
	\xbea
	\tilde H(\eta)&:=&\tilde E(\eta) e^{-i\eta\sigma_2/2}\sigma_3
	e^{i\eta\sigma_2/2}=\tilde E(\eta)(\sin\eta\sigma_1+\cos\eta\sigma_3)\;,
	\label{tilde-h}\\
	\tilde E(\eta)&:=&\frac{i\omega'(\eta)}{2\omega(\eta)}\;,~~{\rm and}~~
	\omega':=\frac{d\omega}{d\eta}\;.
	\label{E-eta}
	\xeea
Up to a factor of $i$, the Hamiltonian (\ref{tilde-h}) describes the interaction 
of a spin 1/2 magnetic dipole with a changing magnetic field whose direction 
rotates uniformly in the $x-z$ plane.

As we mentioned in the preceding section for the Class 3 systems 
$H^{(2)}(t)=H(t)$. Hence direct application of the method of the adiabatic 
product expansion does not lead to a solution of the Schr\"odinger equation 
for the Hamiltonian (\ref{H1-sho-0}) or (\ref{tilde-h}). In this case, either one
constructs the modified adiabatic product expansion of the preceding section
or examines the adiabatic series expansion of Ref.~\cite{pra97-1}. The latter
yields  a series expansion for the evolution operator $\tilde U(\eta)$ of the 
Hamiltonian $\tilde H(\eta)$, namely
	\xbea
	\tilde U(\eta)={\cal T}e^{-i\int_0^\eta \tilde H(s)ds}&=&
	1-i\int_0^\eta  \tilde H(s)ds+\frac{(-i)^2}{2} \int_0^\eta\int_0^\eta 
	{\cal T}[\tilde H(s_1)\tilde H(s_2)]ds_1ds_2+\cdots+\xnn\\
	&&\hspace{2mm}\frac{(-i)^n}{n!}\int_0^\eta\cdots\int_0^\eta 
	{\cal T}[\tilde H(s_1) \cdots \tilde H(s_n)] ds_1\cdots ds_n
	+\cdots\;,
	\label{dyson}
	\xeea
where ${\cal T}$ stands for the time-ordering operator. Since $\tilde H(\eta)$ 
is proportional to $\omega'(\eta)$, for slowly varying $\omega$ one obtains 
an approximate expression for $\tilde U(\eta)$ by computing a finite number 
of terms in this series. This is in fact another generalization of the adiabatic
approximation, because if one keeps only the first term in this series and
neglects the other terms one is essentially neglecting $\tilde H$ or alternatively
$H^{(1)}$. As we explained above, this is just the adiabatic approximation.
If one keeps more terms in this series, then one obtains a better approximation
than the adiabatic approximation. 

\section*{VI. Quadrupole Interaction of  a Spin 1 Particle 
with a Changing Electric Field}

Consider a spin 1 particle  interacting with a changing electric field
$\vec{\cal E}(t)=({\cal E}_1(t),{\cal E}_2(t),{\cal E}_3(t))$ according to 
the Stark Hamiltonian
	\xbe
	H(t)=\lambda[\vec J\cdot\vec{\cal E}(t)]^2\;,
	\label{stark}
	\xee
where $\lambda$ is a real coupling constant and $\vec J$ is the angular
momentum of the particle. The quadrupole interactions of the form
(\ref{stark}) have been extensively studied for fermionic systems in 
relation with the non-Abelian geometric phases \cite{mead,avron1,avron2}
(See also \cite{j-a}.) The occurrence of non-Abelian geometric phases for the degenerate
spin ~1 systems has been pointed out in Ref.~\cite{jpa97}. For these systems, 
the particle has a definite angular momentum $j=1$ and the Hamiltonian is
a $3\times 3$ matrix. Using the spin $j=1$ representation of  $J_i$,
we can express the Stark Hamiltonin (\ref{stark}) in the form
	\xbe
	H=(\frac{\lambda r^2}{2}) \left(\begin{array}{ccc}
	      1+2z^2 & \sqrt{2} z e^{-i\theta} & e^{-2i\theta} \\
	      \sqrt{2} z e^{i\theta} & 2 & -\sqrt{2} z e^{-i\theta} \\ 
	      e^{2i\theta} &-\sqrt{2} z e^{i\theta}& 1+2z^2
	      \end{array}\right)\;,
	\label{stark-matrix}
	\xee
where $r,~\theta,$ and $z$ are defined by
	\[r:=\sqrt{{\cal E}_1^2+{\cal E}_2^2}\;,~~~e^{i\theta}:=
	\frac{{\cal E}_1+i{\cal E}_2}{r}\;,~~
	{\rm and}~~z:=\frac{{\cal E}_3}{r}\;.\]

In view of the general results of Ref.~\cite{jpa97}, if $r\neq 0$ then the
Hamiltonian (\ref{stark-matrix}) has a degenerate and a nondegenerate 
eigenvalue. In the following we shall consider the case where ${\cal E}_3=0$. 
The general case ${\cal E}_3\neq 0$ can be similarly treated. 

If ${\cal E}_3=0$, then $z=0$ and
	\xbe
	H=(\frac{\lambda r^2}{2}) \left(\begin{array}{ccc}
	      1 & 0 & e^{-2i\theta} \\
	     0 & 2 & 0 \\ 
	      e^{2i\theta} & 0 & 1
	      \end{array}\right)\;.
	\label{stark-matrix-0}
	\xee
The eigenvalues of this Hamiltonian are given by
	\xbe
	E_1=0,~~~~E_2=\lambda r^2\;.
	\label{S-eg-va}
	\xee
For $r\neq 0$, $E_1$ is nondegenerate and $E_2$ is doubly degenerate.
A set of orthonormal eigenvectors of this Hamiltonian is given by
	\xbe
	|\psi_1;R\xkt:=\frac{1}{\sqrt{2}}\left(\begin{array}{c}
	-1\\ 0\\e^{2i\theta}\end{array}\right)\;,~~~
	|\psi_2,1;R\xkt:=\frac{1}{\sqrt{2}}\left(\begin{array}{c}
	1\\ 0\\e^{2i\theta}\end{array}\right)\;,~~~
	|\psi_2,2;R\xkt:=\left(\begin{array}{c}
	0\\ 1\\0\end{array}\right)\;.
	\label{S-eg-ve}
	\xee
where $R=(r,\theta)$.

Next we compute $U^{(0)}(t)$ for this system. In order to do this
we first use Eq.~(\ref{a=}) to calculate ${\cal A}^n$. In view of the
fact that the Hamiltonian (\ref{stark-matrix-0}) is Hermitian, 
$|\phi_n,a;R\xkt=|\psi_n,a;R\xkt$ and Eq.~(\ref{a=}) leads to
	\xbe
	{\cal A}^1=-d\theta\;,~~~{\cal A}^2=\left(\begin{array}{cc}
	-d\theta&0\\
	0&0\end{array}\right)\;.
	\label{S-a=}
	\xee
Substituting Eqs.~(\ref{S-a=}) into Eq.~(\ref{6}) and making use of 
Eq.~(\ref{S-eg-va}), we find
	\xbe
	K^1(t)=e^{-i[\theta(t)-\theta_0]} \;,~~~~
	K^2(t)=e^{-i\rho(t)}\left(\begin{array}{cc}
	e^{-i[\theta(t)-\theta_0]}&0\\
	0&1\end{array}\right)\;,
	\label{K's}
	\xee
where 
	\xbe 
	\theta_0:=\theta(0)\;,~~{\rm and}~~\rho(t):=
	\lambda\int_0^t r(s)^2ds\;.
	\label{theta-rho}
	\xee
Using Eqs.~(\ref{u0}) and (\ref{S-eg-ve}), we have
	\xbe
	U^{(0)}(t)=\frac{1}{2}\left(\begin{array}{ccc}
(1+e^{-i\rho(t)})e^{i\theta_-(t)}&0&(-1+e^{-i\rho(t)})e^{-i\theta_+(t)}\\
0 & 2e^{-i\rho(t)} & 0\\
(-1+e^{-i\rho(t)})e^{i\theta_+(t)}&0&(1+e^{-i\rho(t)})e^{i\theta_-(t)}
	\end{array}\right)\;,
	\label{S-u0}
	\xee
where $\theta_\pm(t):=\theta(t)\pm\theta_0$.

Next we compute the Hamiltonian $H^{(1)}(t)$. This involves the 
calculation of  $A_{ab}^{nm}(t)$ and $H_{ab}^{mn}(t)$ for $m\neq n$. 
Using Eqs.~(\ref{H1-ab}) and (\ref{S-eg-ve}) we have
	\xbea
	A^{12}_{11}&=&A^{21}_{11}=-\dot\theta\;,~~~
	A^{12}_{12}=A^{21}_{21}=0\;,\xnn\\
	H^{21}_{11}&=&H^{12}_{11}=\dot\theta e^{i\rho(t)}\;,~~~
	H^{12}_{12}=H^{21}_{21}=0\;.\xnn
	\xeea
Substituting these equations in Eq.~(\ref{H1}) and using 
Eqs.~(\ref{S-eg-ve}), we obtain
	\xbe
	H^{(1)}(t)=-\dot\theta(t) \left(\begin{array}{ccc}
	\cos\rho(t)&0&-i\sin\rho(t) \\
	0& 0& 0\\
	i\sin\rho(t) &0&-\cos\rho(t)
	\end{array}\right)=-\dot\theta(t)[\sin\rho(t)\Sigma_2+
	\cos\rho(t)\Sigma_3]\;,
	\label{S-H1}
	\xee
where
	\xbe
	\Sigma_2:=\left(\begin{array}{ccc}
	0&0&-i\\
	0&0&0\\
	i&0&0\end{array}\right) \;,~~{\rm and}~~
	\Sigma_3:=\left(\begin{array}{ccc}
	1&0&0\\
	0&0&0\\
	0&0&-1\end{array}\right)\;.
	\label{sigma23}
	\xee
It is not difficult to recognize $\Sigma_2$ and $\Sigma_3$ as the Pauli
matrices $\sigma_2$ and $\sigma_3$ represented in a $(0+1/2)$ 
representation of $SU(2)$. In view of this identification we can express
$H^{(1)}(t)$ in the form
	\xbe
	H^{(1)}(t)=-\dot\theta(t) e^{i\rho(t)\Sigma_1/2}\Sigma_3\;
	e^{-i\rho(t)\Sigma_1/2}\;,
	\label{S-H1-su2}
	\xee
where
	\[\Sigma_1:=\left(\begin{array}{ccc}
	0&0&1\\
	0&0&0\\
	1&0&0\end{array}\right)\;,\]
and we have used Eq.~(\ref{identity}).

The Hamiltonian $H^{(1)}(t)$ has the following interesting properties.
	\begin{itemize}
	\item[a)] For $\dot\theta=0$, i.e., $\theta=$ constant, the adiabatic
	approximation is exact and $U(t)=U^{(0)}(t)$.
	\item[b)] In view of  Eq.~(\ref{S-H1-su2}) for $\dot\theta\neq 0$,
	$H^{(1)}(t)$ has three nondegenerate eigenvalues, namely
	$-\dot\theta,~0,$ and $\dot\theta$. This is quite remarkable
	because it shows that the adiabatic canonical transformation
	maps the degenerate Hamiltonian (\ref{stark-matrix-0}) into
	the nondegenerate Hamiltonian (\ref{S-H1-su2}).
	\item[c)] Since $\Sigma_i$ is a representation of the Pauli matrix
	$\sigma_i$, the Hamiltonian (\ref{S-H1-su2}) belongs to a
	representation of the Lie algebra of $SU(2)$. This means that
	one can reduce the Schr\"odinger equation for this Hamiltonian
	to that of the dipole Hamiltonian  \cite{pra97-1,jmp97-2}
	\[H_{\rm dp}=-2\dot\theta(t) e^{i\rho(t)J_1}J_3e^{-i\rho(t)J_1}\;.\]
	\item[d)] We can perform another canonical transformation, namely
	the one defined by $g(t)=\exp[-i\rho(t)\Sigma_1/2]$ to transform
	the Hamiltonian (\ref{S-H1-su2}) into
	\xbe
	H^{(1)'}(t)=\frac{r(t)^2}{2}\Sigma_1-\dot\theta(t)\Sigma_3\;,
	\label{S-H1-prime}
	\xee
	where we have used Eqs.~(\ref{S-H1-su2}), (\ref{H-trans}) and 
	(\ref{theta-rho}).
	In particular if $\dot\theta$ and $r^2$ happen to be proportional,
	i.e., for some $c\in\xR$
	\xbe
	\dot\theta(t)=c\;r(t)^2\;,
	\label{condi-ex}
	\xee
	then $H^{(1)'}(t) =r(t)^2(\frac{1}{2}\Sigma_1-c\Sigma_3)$. In this
	case the eigenvectors of $H^{(1)'}(t)$ are constant and the adiabatic
	approximation yields the exact solution of the Schr\"odinger equation
	for $H^{(1)'}(t)$. The corresponding evolution operator is then
	given by 
	\xbe
	U^{(1)'}(t) =e^{-i(\frac{1}{2}\Sigma_1-c\Sigma_3)\int_0^t r(s)^2ds}\;.
	\label{u1-prime}
	\xee
	Having obtained the evolution operator for $H^{(1)'}(t)$ we can
	use Eq.~(\ref{u-trans}) to obtain the evolution operator for
	$H^{(1)}(t)$ and $H(t)$. This yields the following expression for
	the evolution operator for $H(t)$:
	\xbe
	U(t)=U^{(0)}(t) e^{i\rho(t)\Sigma_1/2} U^{(1)'}(t) \;,
	\label{u=ex}
	\xee
	where $U^{(0)}(t)$ and $U^{(1)'}(t)$ are given by 
	Eqs.~(\ref{S-u0}) and (\ref{u1-prime}).
	\end{itemize}
The above analysis shows that the condition~(\ref{condi-ex}) defines
a class of exactly solvable time-dependent Stark Hamiltonians. If
$\theta=\omega t$, for some constant frequency $\omega$, this condition
corresponds to the case of the rotating electric field $\vec E=r(
\sin\omega t,\cos\omega t,0)$ with magnitude $r$.

\section*{VII. Conclusion}

In this article we have extended the method of the adiabatic product
expansion to non-Hermitian and degenerate Hamiltonians. We showed that in 
general there were three possibilities for the adiabatic product expansion:
	\begin{itemize}
	\item[1)] The expansion terminates after a finite number of 
iterations. This happens when one of the transformed Hamiltonians 
vanishes. In this case the method yields the exact solution for the 
Schr\"odinger equation;
	\item[2)] The expansion consists of an infinite number of distinct
terms. In this case, the method does not lead to an exact solution, but
it gives rise to a generalization of the adiabatic approximation. This
approximation is performed by keeping a finite number of terms in
the product expansion. The general asymptotic behaviour of the adiabatic
product expansion has not been studied. However, one can interpret
this approximation by recalling that the condition for the termination
of the product expansion corresponds to the validity of the conventional
adiabatic approximation for one of the transformed Hamiltonians.
	\item[3)] The expansion involves terms which are not distinct. In this
case the expansion does not lead to a solution. However, usually one can
make another time-dependent canonical transformation after each adiabatic
transformation and obtain an infinite product expansion with the properties
of case 2) above. 
	\end{itemize}

We have considered some specific problems that one can attempt
to solve using this method. We treated the case of a general nondegenerate
two-level system and applied our general results to the more specific
case of the classical equation of motion for a harmonic oscillator with
a time-dependent frequency. In this case, we showed that the adiabatic
canonical transformation mapped the corresponding two-level quantum
system to a quantum system with an anti-Hermitian Hamiltonian. Although
the direct application of the method of adiabatic product expansion did
not yield a solution, we could construct the modified adiabatic product 
expansion.  We have also outlined an adiabatic series expansion for the 
time-evolution operator of this system which led to another generalization of 
the adiabatic approximation. Finally, we considered the application of our 
method to treat the quadrupole interaction of a spin~1 particle with a changing
electric field. The corresponding (Stark) Hamiltonian had a nondegenerate
as well as a degerenrate eigenvalue. We showed that the adiabatic
canonical transformation mapped this Hamiltonian to a Hamiltonian
which had nondegenerate eigenvalues and belonged to a reducible
$(0+1/2)$ representation of the Lie algebra of $SU(2)$. This
means that we can directly use the results of Refs.~\cite{pra97-1,jmp97-2}
which treat the Schr\"odinger equation for a nondegenerate Hamiltonian
belonging to (an irreducible representation of) the Lie algebra of $SU(2)$.
Furthermore, we identified a class of exactly solvable spin 1 quadruple 
Hamiltonians.

\end{document}